\documentstyle[aps,prl,amsfonts]{revtex}
\input epsf
\begin{document}
\draft
\twocolumn[\hsize\textwidth\columnwidth\hsize\csname@twocolumnfalse%
\endcsname
\title{On the Microscopic Origin of Cholesteric Pitch}
\author{A.B.~Harris, Randall D.~Kamien and T.C.~Lubensky}
\address{Department of Physics and Astronomy, University of Pennsylvania,
Philadelphia, PA 19104}

\date{\today}
\maketitle
\begin{abstract}
We present a microscopic analysis of the instability of the
nematic phase to chirality when molecular chirality
is introduced perturbatively.  We show that for central force
interactions the previously neglected
short--range biaxial correlations play a crucial role
in determining the cholesteric pitch. We propose a pseudoscalar
strength which quantifies the chirality of a molecule.
\end{abstract}
]

\begin{figure}
\centerline{
\epsfxsize=2.5truein
\epsfbox{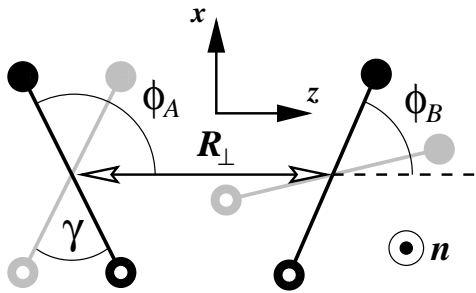}}
\vskip 10pt
\caption{Schematic representation of two chiral molecules.  The atoms are
represented by both the filled and unfilled circles, while the lines serve
only as a guide to the eye.  Each line on each molecule lies in an $xz$
plane parallel to the page.  The arms and atoms in the plane at $y=L/2$
are black and those in the plane at $y=-L/2$ are grey.
The angle $\gamma$ between the projection of the two arms onto the
same $xz$ plane determines the degree of chirality.  As examples,
we consider two versions of this molecule.  In the first, all atoms are
identical, while in the second, the atoms with a hollow center carry a
negative charge and those with a filled center carry a positive charge.  In
the nematic phase, the molecules spin freely about the nematic axis normal
to the page so that $\langle\sin\phi_{\scriptscriptstyle X}
\rangle=\langle\cos\phi_{\scriptscriptstyle X}\rangle =0$.
There are, however, orientational correlations between 
$\phi_{\scriptscriptstyle A}$ and
$\phi_{\scriptscriptstyle B}$.}
\label{fig1}
\end{figure}

Chirality in molecules leads to a myriad of macroscopic chiral structures,
including
life itself \cite{Pasteur}.  A molecule is chiral if its mirror image cannot
be rotated to replicate itself \cite{Kelvin}.  Equivalently it is chiral if its
symmetry
group does not contain
the element ${\bf S}_n$ -- a rotation around an $S_n$ axis by $2\pi/n$ followed
by a mirror
through a plane perpendicular to that axis.  
A chiral molecule cannot be uniaxial: the only 
infinite point groups are ${\cal C}_{\infty v}$ and ${\cal D}_{\infty h}$
\cite{Landau} which both contain ${\bf S}_1$.  
Thus if the molecular orientations are
averaged independently (even if the distribution is
uniaxial about a common molecular axis)
the interactions will be identical to those of molecules
with a $C_\infty$ axis and will therefore not be chiral.
Figure
\ref{fig1} shows a schematic representation of two interacting chiral
molecules whose degree of chirality can be varied continuously as we will
discuss below.
Molecular chirality induces chiral interactions that
produce intermolecular torques of a given sign and can give rise to 
equilibrium chiral structures such as the the cholesteric phase of liquid
crystals. There are two common analyses of this effect. The first is 
purely classical\cite{Straley,Evans,Vertogen,Pelcovits,Moro}, while the
second invokes a generalized chiral dispersion force whose origin is quantum
mechanical\cite{Vertogen}.  
One can argue that there are systems for which the 
dominant interaction is a classical one, involving 
two-point, central forces between atoms or interaction centers on 
molecules, and accordingly
in this paper we consider only this classical mechanism.
In a more detailed paper\cite{longpaper}, we will consider the quantum
interaction and compare its strength with the classical one studied here.  Our
primary focus will be the calculation of cholesteric pitch from these
interactions. The usual classical picture of the origin of intermolecular
twist considers two screw-like molecules with excluded volume
interactions\cite{Straley,Evans}. In order for the vanes of  the screws to
interleave, the molecules must have a non-zero angle between their screw
axes.  A similar picture arises via the tangent-tangent interactions of chiral
molecules\cite{Vertogen,Pelcovits} or via surface-nematic interactions of
chiral dopants\cite{Moro}. 
\par
The above mechanism produces 
a preferred nonzero rotation angle between long-axes of
neighboring chiral molecules (such as two screws) only if there
is correlation between the directions of their short axes, e.g. between
$\phi_{\scriptscriptstyle A}$ and $\phi_{\scriptscriptstyle B}$
in Figure \ref{fig1}.   To our knowledge, no previous attempt to calculate
the cholesteric pitch has treated this properly. If neighboring
molecules are spun freely about their long axes, they become effectively
nonchiral, and interactions favoring twist are washed out \cite{Salem}.  
Thus, the
pitch of a cholesteric depends critically on the degree of intermolecular
correlation of short axes directions: a vanishing 
correlation leads to an infinite
pitch, and complete correlation, as would be produced by long-range
biaxial order, leads to the shortest pitch.  In a uniaxial phase, 
mean-field theory does not
treat these short-axes correlations and cannot predict a finite
cholesteric pitch
from molecular shape. Thus it will either lead to a phase with long-range
order in the short axes directions (a biaxial phase) or to a uniaxial phase
without chirality\cite{Schroder}. Although the results presented 
here are only methodological, they
have crucial implications for numerical simulations\cite{Stroobants}
since such calculations are often based on excluded-volume
hard-core or other central classical interactions of the type we consider.
In what follows, we will focus on molecules with biaxial symmetry and biaxial
correlations, though our results can be generalized to molecules
with $n$-axial symmetry.  
Cholesteric phases are, of course, biaxial.  We are not 
discussing the higher order effect due to the 
biaxial order induced by the cholesteric pitch axis, but rather,
the reverse mechanism in which biaxial order is needed to 
produce chiral phases.  
   
The physics of many chiral liquid crystalline phases can be captured via the
phenomenological Frank free energy density for the unit vector ${\bf n}$:
\begin{eqnarray}
f= &&\case{1}{2}K_1\left(\bbox{\nabla}\!\cdot\!{\bf n}\right)^2
+\case{1}{2}K_2\left[{\bf n}\!\cdot\!\left(\bbox{\nabla}\!\times\!{\bf n}
\right)\right]^2\nonumber\\
&&+\case{1}{2}K_3\left[{\bf n}\!\times\!\left(\bbox{\nabla}\!\times\!{\bf n}
\right)\right]^2
+ h{\bf n}\!\cdot\!\left(\bbox{\nabla}\!\times\!{\bf n}\right) ,
\label{Frank}
\end{eqnarray}
where $K_1$, $K_2$ and $K_3$ are the Frank elastic constants for splay, twist
and bend, respectively, and $h$ is a chiral parameter.  This free energy
is invariant under the inversion ${\bf n}\rightarrow -{\bf n}$,
consistent with the symmetry of the nematic phase.
The calculation of $h$ is the focus of this paper.

The most common manifestation of the preferred packing angle of chiral
molecules is the cholesteric liquid crystal phase in which a particular axis
of each molecule lies along the nematic director ${\bf n} = \left[\cos kz,
\sin kz,0\right]$.  The $\hat z$ axis is the pitch axis and the pitch is
$P=2\pi/k$, the distance over which the nematic director rotates by $2\pi$.
In this uniform twisted state
${\bf n} \cdot\!\bbox{\nabla}\!\times\!{\bf n} = - k$, and the Frank
free energy density reduces to $f = \case{1}{2} K_2 k^2 - h k$ so that
$ h = - \partial f /\partial k |_{k = 0}$. The equilibrium value
of $k$ is $k_0=h/K_2$. Using standard
statistical mechanical procedures, the chiral parameter can be expressed as
\begin{equation}
h = - \left.{\partial f \over \partial k}\right\vert_{k=0} = - \left.{
1 \over\Upsilon}\left\langle
{\partial U\over \partial k}\right\rangle\right\vert_{k=0},
\label{h-U}
\end{equation}
where the brackets denote a
thermodynamic average in a nematic state in which there is a spatially
uniform director ${\bf n}$, $\Upsilon$ is the sample volume, and
$U$ is the total potential energy.
We denote the center-of-mass coordinate
of molecule $A$ by ${\bf R}_{\scriptscriptstyle A}$ 
and the coordinate relative to the center
of mass of particle $\alpha$ in molecule $A$ by ${\bf r}_{{\scriptscriptstyle
A}\alpha}$.  In
general, $\alpha$ should run over all interaction centers (usually atoms
and nuclei) in molecule $A$.  The potential energy
$U$ is a function of all the coordinates. To
determine $h$, we introduce an infinitesimal twist in a nematic state in which
the molecules are aligned along a uniform director
${\bf n}$. Under such a twist, atomic coordinates within molecule $A$ will
undergo
a rotation $\delta r_{{\scriptscriptstyle A}\alpha}^i = \epsilon_{ijk} \delta \omega_{\scriptscriptstyle A}^j
r_{{\scriptscriptstyle A}\alpha}^k$ 
where $\delta\omega_{\scriptscriptstyle A}^j = 
k e^j ({\bf e}\cdot {\bf R}_{\scriptscriptstyle A})$ is a
rotation
angle about an arbitrary unit vector ${\bf e}$ perpendicular to ${\bf n}$. The
magnitude of
$\delta \omega_{\scriptscriptstyle A}$ increases linearly with the projection of the
center-of-mass position ${\bf R}_{\scriptscriptstyle A}$ along ${\bf e}$.  Using (\ref{h-U}) and
the invariance of the system with respect to arbitrary rotations about the
${\bf n}$-axis, we obtain
$h = - {1 \over 4}\Upsilon^{-1} \sum_{\scriptscriptstyle BA} T_{\scriptscriptstyle BA}$,
where $T_{\scriptscriptstyle BA}=\langle{\bf R}_{\perp}
\!\cdot\! \bbox{\tau}_{\!\scriptscriptstyle BA}\rangle$ is the projected
torque,
$\bbox{\tau}_{\!\scriptscriptstyle BA}$ is the torque exerted on molecule $B$
by molecule $A$,
${\bf R} = {\bf R}_{\scriptscriptstyle B} - {\bf R}_{\scriptscriptstyle A}$, and ${\bf R}_{\perp}$ is the projection of
${\bf R}$ onto the plane perpendicular to ${\bf n}$.

The chiral parameter $h$ must be zero in achiral systems that have an
equilibrium nematic phase.  To see that our formulation leads to this result,
consider
evaluating $\sum_{\scriptscriptstyle A} T_{\scriptscriptstyle BA}$ for a system in which all
molecules are achiral.  The thermodynamic average is carried out over all
configurations
$\bbox{\Omega}$ of the molecules consistent with the assumed nematic order.
If, and
only if, the molecules are achiral, the average may equivalently be carried out
over all configurations $\overline{\bbox{\Omega}}$, where
$\overline{\bbox{\Omega}}$ is obtained
from $\bbox{\Omega}$ by a reflection through a plane perpendicular
to $\bf n$.  But
${\bf R}_{\perp}\!\cdot\bbox{\tau}_{\scriptscriptstyle BA}$ in
$\bbox{\Omega}$ is
the negative of ${\bf R}_{\perp}\!\cdot\bbox{\tau}_{\scriptscriptstyle BA}$
in $\overline{\bbox{\Omega}}$, so $h=-h=0$.

Our expression for $h$ is perfectly general: it applies to quantum as well
as classical systems.  Note that hard-core interactions
can be viewed as the limiting case of central forces between
atoms of a single kind that mutually interact via central forces, and
thus we begin our analysis for such systems.
The projected torque is then
\begin{equation}
T_{\scriptscriptstyle BA} = \left\langle\sum_{\beta\alpha}
\epsilon_{ijk}R_{\perp}^i r_{\beta}^j
\partial^kV({\bf R} + {\bf r}_{\beta} - {\bf r}_{\alpha} ) \right\rangle
\label{Tabcl}
\end{equation}
where ${\bf r}_{\alpha} = {\bf r}_{{\scriptscriptstyle A}\alpha}$ and ${\bf r}_{\beta} = {\bf
r}_{{\scriptscriptstyle B}\beta}$ are,
respectively, the coordinates of atoms $\alpha$ and $\beta$ in molecules $A$
and $B$. To facilitate our analysis, we will now consider an expansion of
$T_{\scriptscriptstyle BA}$ in powers of relative atomic distance over
center-of-mass separation,
{\sl i.e.}, in $r_{\beta,\alpha}/R$. We expect, however, that the conclusions
we draw
from this analysis are more generally valid and apply, in particular, to 
hard-core interactions.  In such an expansion, only terms that are odd in the
atomic coordinates ${\bf r}_{\alpha}$ and ${\bf r}_{\beta}$ are sensitive to
reflections and thus to chirality.  Furthermore, terms that are even in
${\bf r}_{\alpha}$ and ${\bf r}_{\beta}$ are necessarily odd in ${\bf R}$ and
will,
therefore, not survive the average over the nematic distribution function.
Thus, if we assume an achiral distribution characteristic of a nematic phase,
we can restrict our attention to terms odd in ${\bf r}_{\alpha}$ and
${\bf r}_{\beta}$.  (We will, however, reconsider this point later).
Since ${\bf r}_{\beta}$ measures the position relative to the
center of mass, $\sum_{\beta} {\bf r}_{\beta} = 0$, and thus the linear and
third order terms vanish.  The first non-vanishing
chiral term is the fifth order term
\begin{eqnarray}
&& T_{\scriptscriptstyle BA}^5  = \label{fifthorder}\\
&& \quad{1 \over 4!}\left\langle
\sum_{\beta\alpha}\epsilon_{ijk}R_{\perp}^i
r_\beta^jr_{\beta\alpha}^lr_{\beta\alpha}^m
r_{\beta\alpha}^pr_{\beta\alpha}^s
\partial^k\partial^l\partial^m\partial^p\partial^sV({\bf R})\right\rangle ,
\nonumber
\end{eqnarray}
where ${\bf r}_{\beta\alpha} = {\bf r}_\beta -{\bf r}_\alpha$.
This quantity can be reexpressed in terms of the second and third rank
mass moment tensors on molecules $X=A,B$,
$M_X^{jl} = \sum_{\chi\in X}( r_{\chi}^j r_{\chi}^l - \case{1}{3} r_{\chi}^2
\delta^{jl})$ and $S_X^{jlm} = \sum_{\chi\in X}r_{\chi}^j r_{\chi}^l
r_{\chi}^m$.  $M^{jl}$ is the usual quadrupole tensor
describing nematic order, which can be decomposed into a uniaxial and a
biaxial part.  If we assume that there is perfect alignment of the longest
principal axis along the
nematic direction, then $M^{jl} = M_1 ( n^j n^l -\case{1}{3} \delta^{jl}) +
B^{jl}$, where $B^{jl}$ is the symmetric traceless biaxial tensor with no
components along ${\bf n}$.  The third rank tensor
$S^{ijk}$ is symmetric.
In general, there are correlations between the direction of the vector
${\bf R}$ connecting two molecules and the respective orientations of these
molecules.  The important effects we consider are present even if
these correlations are absent, and we will ignore them.  This permits us to
evaluate the orientational average of products of $R^i$ with respect to a
distribution that is isotropic in the plane perpendicular to ${\bf n}$.  for
example $\langle R^i R^j \rangle = R_{||}^2 n^i n^j + \case{1}{2}R_{\perp}^2
(\delta^{ij} - n^i n^j)$. Setting
$V({\bf R} ) = g(R^2/2)$, expanding the derivatives of $V$, and performing the
above average, we obtain
\begin{equation}
T_{\scriptscriptstyle BA}^5 = \case{1}{2}
\epsilon_{ijk}Q^{ip}\left\langle\left( B_{\scriptscriptstyle B}^{jl}
S_{\scriptscriptstyle A}^{kpl} +
B_{\scriptscriptstyle A}^{jl} 
S_{\scriptscriptstyle B}^{kpl} \right) K({\bf R})\right\rangle ,
\label{TAB5}
\end{equation}
where $Q^{ip} = n^i n^p - \case{1}{3}\delta^{ip}$ and
\begin{equation}
K({\bf R})= R_{\perp}^2\Big\{g^{(3)} + R_{||}^2 g^{(4)} +
\case{1}{4}R_{\perp}^2
\big[g^{(4)}+R_{||}^2 g^{(5)}\big]\Big\},
\end{equation}
where $g^{(n)}(x) = d^n g(x)/dx^n$.  We see then, that only the
traceless part of $S^{kpl}$ contributes to $T^5$ and so we may
take it to be traceless.

As we have already discussed, $T_{\scriptscriptstyle BA}$ is nonzero only if
the molecules are
chiral.  How is this fact manifested in (\ref{TAB5})? Since both
$Q^{ip}$ and $B^{jl}$ can be nonzero for achiral molecules, it would seem
that the tensor $S^{kpl}$ is a measure of chirality.  This is not true,
however, because $S^{kpl}$ also has components that can be nonzero for
achiral molecules.  Though there are many possible definitions of a
{\it molecular} chiral strength, when a molecule has a unique long axis
we propose the pseudoscalar $\psi = S^{klm}\epsilon_{ijk}Q^{il}B^{jm}$
as a measure of the chiral strength of a molecule.
This definition is useful because, as we shall see, $h$ is proportional to 
$\psi$.  If
the molecule is not biaxial ({\it i.e.}, $B^{km}=0$), $\psi$ will
vanish.  In addition, since $\psi$ is a rotational invariant
{\it odd} in ${\bf r}$,
it will also vanish if the molecule is achiral.  It is possible
that even for a chiral molecule, $\psi$ vanishes.  If this were the case,
however, there would still be nonvanishing contributions to (\ref{Tabcl}) at
higher order in the expansion in powers of $r/R$.  Indeed, a complete
description of chiral interactions requires the knowledge of {\it all} the
chiral moments of the molecules.
In
the basis of the principal axes of the molecule with $\bf n$ along $\hat z$,
$\psi =
S^{xyz}(B^{xx} - B^{yy})$. Only the components of $S^{kpl}$
with three different indices
in this basis contribute to $\psi$.  We can, therefore, replace
$S^{kpl}$ in (\ref{TAB5}) with the tensor
${\overline S}^{kpl}$ whose only nonvanishing component in the principal axis
basis is
$S^{xyz}$.

The projected torque $T_{\scriptscriptstyle BA}^5$ is an average over
fluctuations in the
aligned nematic phase.  It will be zero unless biaxial directions on pairs of
molecules are correlated.  If we were to spin
the molecules independently, then $\langle B^{km}\rangle$ would be zero, and
both terms in (\ref{TAB5}) would vanish.
However, when there are biaxial correlations, this
term will not vanish.  To see this explicitly, we can use the identity
${\overline S}_{\scriptscriptstyle A}^{kpl} =
-\psi_{\scriptscriptstyle A}\epsilon_{prs}
B_{\scriptscriptstyle A}^{kr}Q^{sl}/2B^2 + (\hbox{5 symmetric permutations})$,
where $B^2={\rm Tr}\left(B^2\right)$.  Inserting
this into (\ref{TAB5}), we come to (assuming identical molecules):
\begin{equation}
T_{\scriptscriptstyle BA} =  {\psi \over B^2} \left\langle
K({\bf R}) B_{\scriptscriptstyle A}^{ij}
B_{\scriptscriptstyle B}^{ij}\right\rangle\equiv
 \psi \left\langle K({\bf R}) \Gamma_b ({\bf R})\right\rangle
\label{fti}
\end{equation}
where $\Gamma_b({\bf R})\propto\langle\cos\left[2\left(
\phi_{\scriptscriptstyle B}-\phi_{\scriptscriptstyle A}\right)
\right]\rangle$ is the
biaxial correlation function for two molecules separated by a distance $R$.
We would expect in a uniaxial phase that $\Gamma_b(R) \propto e^{-R/\xi}$ where
$\xi$ is the biaxial correlation length.
Naively, one would expect $\xi$ to be of the order of the
molecular spacing.
Thus we conclude that at the very least chirality requires biaxial correlations
among the nematogens.  We tabulate $\psi$ for a number of chiral molecules in
Table \ref{tab1}.
\vbox{
\begin{table}
\caption{Value of $\psi$ for molecules made of atoms located at
the coordinates given in the first column. The first molecule is
shown in Fig.~1 and the second is a helix of uniform density.}
\label{tab1}
\begin{tabular}{lc}
\multicolumn{1}{c}{Atomic Coordinates} & \multicolumn{1}{c}{$\psi$}\\
\hline
\begin{tabular}{l}$L>2w$:\quad$\{(\pm w,0,-L/2),$\\
\qquad$(\pm w\cos\gamma,\pm w\sin\gamma,L/2)\}$
\end{tabular} &
$2w^4L\sin(2\gamma)$\\
&\\
\begin{tabular}{l}
$L\gg r$, $n\in{\Bbb Z}$:\quad$\big\{s\in[-1/2,1/2],$\\
\quad$\big(r\cos(2\pi ns),r\sin(2\pi ns),Ls\big)\big\},$
\end{tabular} &
$\sim
-{\displaystyle 3r^4L\over\displaystyle(2\pi n)^3}\left[1-{\displaystyle
24\over
\displaystyle(\pi n)^2}\right]$
\end{tabular}
\end{table}
}

Equation (\ref{fti}) gives the dominant contribution to $T_{\scriptscriptstyle
BA}$
to linear
order in $\psi$ in the nematic phase.  There is an additional contribution
linear in $\psi$ arising from the chiral part of the equilibrium probability
distribution ($e^{-U/k_{\scriptscriptstyle B}T}/{\cal Z}$ where ${\cal Z}$
is the partition function) and those terms with even powers of $\bf r$ arising
in
$r_\beta^j\partial_kV$ that are averaged in (\ref{Tabcl}).  In the
isotropic phase, our analysis can be extended to show that
this term cancels the contribution
to $T_{\scriptscriptstyle BA}$ from
(\ref{fifthorder}) and (\ref{TAB5}) to produce $h=0$ as required.  In the
ordered
phase, this other term is {\it higher order} in correlation functions and
is subdominant to our result \cite{longpaper}.

We have shown that the projected torque $T_{\scriptscriptstyle BA}$ and hence
$K_2k_0$ will be proportional to the molecular chiral strength $\psi$.
We note that there are a number of chiral liquid crystals,
such as solutions of the viruses FD and TMV as well as of DNA, that
show very small, if any, macroscopic chirality \cite{Fraden,NIH}.
We believe the ideas presented here explain these observations,
although a complete understanding will require a thorough investigation of
the quantum dispersion force.  Helical molecules have very small biaxiality
and hence small values of
$\psi$ (see Table \ref{tab1}). 
The chiral contribution to $\psi$ comes from 
$S^{kpl}$ and is inversely proportional to the number of turns per unit length,
which is consistent with one's geometric
intuition.  In addition, since the molecular chiral strength depends on
the degree of molecular biaxiality, we see that
for {\it fixed} turn density (or equivalently, fixed $S^{kpl}$), $\psi$ 
falls off as the total length {\it squared}.
In addition, these molecules can easily rotate independently
and are far apart (tens of Angstroms).  Hence we believe the
biaxial correlation length $\xi$ will be small compared to the intermolecular
spacing. Our classical
analysis should be valid especially for FD and TMV since these molecules
are thought to interact sterically.
Alternatively, short-molecule, thermotropic liquid crystals
show very strong chirality, with pitches on the order of $5000\AA$.  These
molecules are generally quite flat and thus quite biaxial.
Typical molecular densities would not allow the molecules
to rotate independently of each other, and thus
we expect the biaxial correlations at the molecular separation to be reasonably
large. The combined effect of a large biaxial component to $\psi$ and of a
large $\xi$ should lead to relatively short pitches. In both cases we note that
naively
one would expect on dimensional grounds
that $k_0$ would be on the order of $\pi/a$ where $a$ is the
intermolecular spacing, which is certainly not a typical inverse pitch.
Pitches are typically on the scale of {\it microns}, not {\it Angstroms}.
Our expression
(\ref{fti}) for the leading term of $T_{\scriptscriptstyle BA}$
is consistent with all of the above observations.  We also note
that in all but the most dilute solutions
we do not expect any universal dependence of pitch on concentration
or temperature: the details of the interactions and correlations should
be different from system to system.

We briefly mention a number of generalizations to be discussed later 
\cite{longpaper}.  We have considered here only
the interactions between a pair of perfectly aligned, identical molecules.
In the nematic state, the molecules are not perfectly aligned and
the Maier-Saupe order parameter $S$ is less than $1$.
We can incorporate these
fluctuations
into the calculation of $T_{\scriptscriptstyle BA}$.  Indeed, we find, as
discussed above, that
when there is no nematic order there
is no net torque.  Since (\ref{TAB5}) involves the product of $S^{kpl}$ on
one molecule and $B^{jl}$ on the other, our results easily generalize to chiral
molecules interacting with achiral, biaxial molecules.
More generally, we find that including
correlations between the intermolecular direction and the molecular
orientation leads to chiral interactions between chiral molecules and
uniaxial
molecules.

Additionally, we note that atomic identity may be relaxed.
In this case, there is an
interaction $V_c$ between pairs of atoms leading to a potential energy
$U =\case{1}{2}\sum q_{\alpha}q_{\beta}V_c
({\bf R}+
{\bf r}_{{\scriptscriptstyle B}\beta}
-{\bf r}_{{\scriptscriptstyle A}\alpha} )$, where 
$q_{\alpha}$ and $q_{\beta}$ are
the
``charges'' of atoms $\alpha$ and $\beta$.  A chiral molecule such as that
shown
in Figure \ref{fig1}\ can carry a dipole moment ${\bf p}_{\scriptscriptstyle A}
 = \sum_{\alpha}
q_{\alpha}{\bf r}_{{\scriptscriptstyle A}\alpha}$ perpendicular to its long axis.
We then find a third order contribution to the projected
torque of the form $\epsilon_{ijk}p_{\scriptscriptstyle A}^l 
C_{\scriptscriptstyle B}^{jm} R_{\perp}^i \partial^k
\partial^l \partial^m V_c$, where $C_B^{jm} = \sum_{\beta}
q_\beta\big(r_{\beta}^j
r_{\beta}^m - \case{1}{3} r_{\beta}^2 \delta^{jm} \big)$ is the charge
quadrupole moment tensor.  The molecular chiral strength analogous to $\psi$ for
this
system is $\psi_c = \epsilon_{ijk}p^iC^{jm} Q^{km}$.  With this definition,
$T_{\scriptscriptstyle BA}$ is proportional to $\psi_c \Gamma_p ( {\bf R} )$
where $\Gamma_p ( {\bf R} )$ is the
dipole-moment pair correlation function, which, like $\Gamma_b ( {\bf R})$
measures
angular correlations in the plane perpendicular to ${\bf n}$.  Adding charges
$q$ and $-q$
to the molecules shown in Figure \ref{fig1} (and using the atomic positions
specified in
Table \ref{tab1}), 
we have $\psi_c = 4q^2w^2L\sin\gamma$.  We believe that chiral
interactions of this type
play a role in those liquid crystals which can form ferroelectric phases ({\it
i.e.},
Sm-$C^*$, ${\rm TGB}_C^*$, {\it etc.}). 

We close with some observations concerning the relation of our work to
previous treatments of chiral interactions.  An intermolecular potential of
the form
$V_{\scriptscriptstyle BA}^{\rm ch} =  M_{\scriptscriptstyle A}^{ij} 
\epsilon_{jkl} R^k M_{\scriptscriptstyle B}^{il} 
V_p ( R) + ( A
\leftrightarrow B )$
remains chiral and nonzero upon spinning about the local nematic director. 
It leads automatically to a free energy of the form of (\ref{Frank}) with $h$
proportional to $V_p$.  Thus, this potential or ones similar to it
are often used as a starting point for the description of chiral liquid
crystals.  Our analysis shows that this potential {\it cannot} be obtained
from classical central forces between atoms on molecules, though
it can arise through quantum dispersion forces \cite{Vertogen}.  The potential
corresponding to (\ref{TAB5}) has the form $V_{\scriptscriptstyle BA}=\sum_{\scriptscriptstyle BA}[S_{\scriptscriptstyle A}^{ijk} M_{\scriptscriptstyle B}^{lm}
\partial^i\partial^j\partial^k\partial^l\partial^m V +(A \leftrightarrow B)]$.
$S^{ijk}$ is a symmetric tensor -- it cannot be expressed in terms
of $\epsilon_{ijk}$ and $M^{kl}$ to produce 
$V_{\scriptscriptstyle BA}^{\rm ch}$.  

It is a pleasure to acknowledge stimulating
communications with D.~Andelman, S.~Fraden, R.~Meyer, M.~Osipov, R.~Pelcovits, R.~Petschek
and
J.~Selinger. ABH was supported by NSF Grant Number 95-20175.  RDK and TCL were
supported by NSF Grant Number DMR94-23114.

\end{document}